\documentclass[]{iopart}
\usepackage{amsbsy}
\usepackage{latexsym}
\usepackage{graphicx}
\usepackage{url}

\begin{document}

\title{Triangulation of gravitational wave sources with a network of
detectors}

\author{Stephen Fairhurst}
\address{Cardiff School of Physics and Astronomy,
Cardiff University, Queens Buildings, The Parade, Cardiff. CF24 3AA
}
\eads{\mailto{Stephen.Fairhurst@astro.cf.ac.uk}}

\begin{abstract} 

There is significant benefit to be gained by pursuing multi-messenger
astronomy with gravitational wave and electromagnetic observations.  In
order to undertake electromagnetic follow-ups of gravitational wave
signals, it will be necessary to accurately localize them in the sky.
Since gravitational wave detectors are not inherently pointing
instruments, localization will occur primarily through triangulation
with a network of detectors.  We investigate the expected timing
accuracy for observed signals and the consequences for localization.  In
addition, we discuss the effect of systematic uncertainties in the
waveform and calibration of the instruments on the localization of
sources.  We provide illustrative results of timing and localization
accuracy as well as systematic effects for coalescing binary waveforms.

\end{abstract}

\section{Introduction}
\label{sec:intro}

There is a growing realization that multi-messenger astronomy will be of
critical importance for gravitational wave astronomy.  While the concept
has been discussed for many years (e.g.  \cite{Schutz:1986gp}), only
recently has a large push towards joint observations with
electromagnetic and neutrino detectors begun (see
e.g.~\cite{Stamatikos:2009ja, Prince:2009uc, Bloom:2009vx,
Phinney:2009ty}).  Initially, joint observations will provide additional
confidence for early detections made in non-stationary gravitational
wave data.  Later, multi-messenger observations will be critical in
extracting the maximum scientific payoff from gravitational wave
observations.  For example, the cleanest way to demonstrate that the
progenitors of short $\gamma$--ray bursts (GRBs) are coalescing Neutron
stars would be the observation of a gravitational wave chirp associated
with a short GRB \cite{nakar07}.  Additionally, joint observations will
likely provide measurements of complementary parameters, thereby
breaking degeneracies which would exist with single messenger
observations.     

The road to joint observations has already been paved, with several
gravitational wave search results being ``triggered'' by external
observations, such as GRBs \cite{Abbott:2008zzb, GRB070201} and soft
gamma repeaters \cite{Abbott:2008gj, Collaboration:2009zd}.  More
recently, work has begun to ensure that gravitational wave observations
can be followed up by other astronomical observatories.  In the era of
regular gravitational wave observations, it is likely that gravitational
wave alerts will be followed up by large field of view optical (such as
Pan Starrs \cite{panstarrs}), $\gamma$--ray (Swift \cite{swift}, Fermi
\cite{fermi}) and radio observatories (Lofar\cite{lofar, Fender:2006qg})
as well as neutrino observatories (Ice Cube \cite{icecube}, Antares
\cite{antares}).

The vast majority of electromagnetic observatories are, by their very
nature, directional.  Thus, in order for gravitational wave observations
to be useful to other astronomers, it is necessary to extract the sky
location from the gravitational wave signal.  However, gravitational
wave detectors are sensitive to signals from large fraction of the sky
and a single gravitational wave detector provides essentially no
directional information for a short duration source.  Thus, the ability
to reconstruct the location of a transient signal is primarily due to
triangulation based on the observed time delays of the signal at several
detectors.  For more than two sites, requiring a consistency between the
observed amplitudes will also serve to restrict the allowed sky
positions.  In particular, for three detectors, using only timing
information, one obtains two sky locations which are mirror images with
respect to the plane of the three detectors; generically the amplitude
information can be used to break this degeneracy.  The issue of
localization of gravitational wave signals with a detector network has
been discussed previously, and several different algorithms proposed
\cite{Dhurandhar:1988my, GurselTinto, Searle:2007uv, Searle:2008ap,
Cavalier:2006rz, 1742-6596-122-1-012038, Beauville:2005au, birindelli,
King:2008, markowitz:122003}.     

In this paper, we consider the ability of gravitational wave detectors
to localize transient signals by considering only the timing information
available at each site.  For concreteness, we restrict attention to
elliptically polarized gravitational waves, a choice motivated by
coalescing binary waveforms.  With this simple model, we obtain a
straightforward estimate for the expected timing accuracy and use this
to evaluate the triangulation ability of a network of detectors.  From
timing information alone, it is only possible to measure the projection
of the sky location onto the detector baseline.  Thus for two detectors,
localization to a ring in the sky is possible, while for three detectors
a reflection degeneracy in the plane of the detectors remains.

In a gravitational wave search, there are several sources of
uncertainties which will affect the localization ability of the search.
First, when performing a matched filter search, there will typically be
several additional parameters in the waveform model (such as the masses
of a compact binary).  These additional parameters will serve to degrade
the localization ability.  In addition, there are likely to be
differences between the physical waveforms and the templates used.
These arise due to errors in the waveform family due to truncation of
analytic expansions or numerical inaccuracies.  While the inaccuracies
of the waveform are independent of the detector, their effect on the
timing accuracy will depend upon detector sensitivity.  Finally, there
are uncertainties in the calibration of the detectors.  These will
result in the reconstructed gravitational wave strain $h(t)$ differing
from the actual gravitational wave signal.  These calibration
uncertainties will have a similar effect to the use of an incorrect
waveform template.  However, the calibration inaccuracies, as well as
the associated timing errors, will be largely independent in the
different detectors.

The layout of the paper is as follows.  In Section \ref{sec:waveform},
we describe the restriction to elliptical polarization and briefly
review the coalescing binary waveform.  In Section \ref{sec:timing} we
obtain the expected timing accuracies, and in Section \ref{sec:location}
present the localization ability of the network.  Finally, in Section
\ref{sec:systematics}, we discuss the systematic uncertainties and their
effect on timing.  Throughout, we provide an illustrative example of
expected results for binary neutron star (BNS) and binary black hole
(BBH) systems.

\section{The waveform model}
\label{sec:waveform}

In this paper, we will focus primarily on waveforms generated during
binary coalescence.  However, much of the framework introduced is
applicable to a broader class of waveforms.  Therefore, we begin by
laying out the minimal set of assumptions that are made on the form of
the gravitational wave, before moving on to describe the waveform for
binary neutron star and black hole coalescences in more detail.

We restrict attention to elliptically polarized waveforms, following a
definition proposed by Sutton and Poprocki \cite{SuttonPoprocki}.
Specifically, a waveform is said to be elliptically polarized if there
exists a polarization frame such that the two polarizations $h_{+}$ and
$h_{\times}$ of the gravitational wave are related by 
\begin{equation}\label{eq:time_elliptic}
  h_{+}(t) = \eta_{+} \alpha(t) \cos \varphi(t) \, , \quad
  h_{\times}(t) = \eta_{\times} \alpha(t) \sin \varphi(t) \, ,
\end{equation}
where $\alpha(t)$ and $\phi(t)$ are the amplitude and phase of the
waveform and $\eta_{+,\times}$ encode the relative amplitudes of the two
polarizations.  We make the additional requirement that the amplitude is
slowly varying (with respect to the phase), i.e. $\dot{\alpha}/\alpha \ll
\dot{\varphi}$.  Then, in Fourier space, the two polarizations are related
by
\begin{equation}\label{eq:elliptic}
  \tilde{h}_{\times}(f) = i \eta \tilde{h}_{+}(f) \, .
\end{equation}
where $\eta$ $\in[-1,1]$ provides the ratio between the two amplitudes,
with unity corresponding to circular polarization and zero to linear
polarization.
  
The gravitational waveform observed at a detector can be expressed as
\begin{equation}\label{eq:h_t}
  h(t) = F_{+}(\theta, \phi, \psi) h_{+}(t; \boldsymbol\xi) + 
         F_{\times}(\theta, \phi, \psi) h_{\times}(t; \boldsymbol\xi) 
\end{equation}
Here, $F_{+}$ and $F_{\times}$ are the well known detector response
functions (see e.g.~\cite{Fairhurst:2007qj}) which depend upon the sky
location ($\theta$, $\phi$) of the system relative to the detector and
the polarization $\psi$.  We use $\boldsymbol\xi$ to denote any
additional parameters upon which the waveform depends.  It follows
straightforwardly from (\ref{eq:elliptic}) and (\ref{eq:h_t}) that the
gravitational waveform observed in a given detector can be expressed
as:%
\footnote{We make use of the two ``phases'' of the waveform $h_{0}$ and
$h_{\frac{\pi}{2}}$ as these arise naturally in the context of
coalescing binaries, as we shall see in Section \ref{ssec:cbc_waveform}}
\begin{equation}\label{eq:h_det}
  h(t) = A_{0} \, h_{0}(t; \boldsymbol\xi)  + 
         A_{\frac{\pi}{2}} \, h_{\frac{\pi}{2}}(t; \boldsymbol\xi) 
\end{equation}
where
\begin{equation}\label{eq:hf_relation}
  \tilde{h}_{\frac{\pi}{2}}(f; \boldsymbol\xi) = 
  i \tilde{h}_{0}(f; \boldsymbol\xi) \, .
\end{equation}
The constants $A_{0}$ and $A_{\frac{\pi}{2}}$ depend upon the location
of the source relative to the detector and the parameter $\eta$
introduced above.  Finally, we write the waveform explicitly in
terms of amplitude and phase as 
\begin{eqnarray}\label{eq:freq_waveform}
  \tilde{h}_{0}(f; \boldsymbol\xi) = A(f; \boldsymbol\xi) 
    e^{i \Phi(f; \boldsymbol\xi)} \quad \textrm{and} \quad 
  \tilde{h}_{\frac{\pi}{2}}(f; \boldsymbol\xi) = i A(f; \boldsymbol\xi) 
    e^{i \Phi(f; \boldsymbol\xi)} \, . 
\end{eqnarray}

\subsection{Waveforms for Coalescing Binaries}
\label{ssec:cbc_waveform}

For concreteness, let us now specialize to the waveform emitted during
binary coalescence, where we neglect the spin of the two components.
Then, the orbital plane will not precess and the two polarizations of
the waveform can be expressed as \cite{Allen:2005fk}
\begin{eqnarray}
  h_{+}(t) &=& \left( \frac{D_{o}}{D} \right) \left[ \begin{array}{r} 
    (1/2) (1 + \cos^2 \iota) \cos 2 \phi_{o} \, 
    h_{0}(t; t_{o}, D_{o}, m_{1}, m_{2}) \\
    -  \cos \iota \sin 2\phi_{o} \, 
    h_{\frac{\pi}{2}}(t; t_{o}, D_{o}, m_{1}, m_{2})
    \end{array} \right] ,  \nonumber \\
  h_{\times}(t) &=& \left( \frac{D_{o}}{D} \right) 
    \left[ \begin{array}{r}
    (1/2) (1 + \cos^2 \iota) \sin 2\phi_{o} \, 
    h_{0}(t; t_{o}, D_{o}, m_{1}, m_{2}) \\
    + \cos \iota \cos 2 \phi_{o} \, 
    h_{\frac{\pi}{2}}(t; t_{o}, D_{o}, m_{1}, m_{2}) 
    \end{array} \right] .
 \label{eq:h_cbc}
\end{eqnarray}
Here, $D$ is the distance at which the signal is located, $D_{o}$ is a
fiducial distance (e.g.~$1$ Mpc), $t_{o}$ and $\phi_{o}$ are a
reference time and phase for the signal (often taken as the coalescence
time and phase), $\iota$ is the inclination angle of the binary relative
to the line of sight and $m_1$ and $m_2$ are the masses of the binary's
components.  The two phases $h_{0}$, $h_{\frac{\pi}{2}}$ are the
waveforms, normalized for a binary at distance $D_{o}$, and depend upon
the masses of the components as well as the coalescence time $t_{o}$.  

The amplitude and phase of the waveform have been calculated to
exquisite accuracy through the post--Newtonian expansion.  Although the
post--Newtonian expressions formally extend to infinite frequency, the
waveform is truncated at a pre-specified frequency, typically the
innermost stable circular orbit (ISCO).  At higher frequencies, the
finite size of the objects will cause the true waveform to differ
significantly from the post-Newtonian expression.  Furthermore, in many
applications, the restricted post--Newtonian approximation is used where
only the leading order amplitude term is used, while the phase is
evaluated to higher post--Newtonian order.  It is only the restricted
post--Newtonian waveform which can be written in the form
(\ref{eq:freq_waveform}).   The amplitude satisfies $A(f) \propto
f^{-7/6}$ while the detailed phasing evolution will depend critically
upon the masses of the system \cite{Blanchet:2002av}.   This waveform is
appropriate for low mass binaries as the merger occurs at a higher
frequency than the sensitive band of the detectors. We will illustrate
the results in the remainder of the paper with numbers appropriate for a
BNS signal with component masses $1.4 M_{\odot}$ and an ISCO frequency
of 1500 Hz.
 
More recently, breakthroughs in numerical relativity have allowed for
the calculation of entire binary black hole merger waveform
\cite{Pretorius:2005gq}.  Work remains ongoing to cover the full mass
and spin parameter space.  However, for non-spinning binaries with
comparable mass components, the waveform is well understood (see, e.g.
\cite{Campanelli:2005dd, Baker:2005vv, Pretorius:2007nq, Husa:2007zz,
Hannam:2009rd}).  Indeed, a phenomenological fit to these waveforms has
been produced in \cite{Ajith:2007qp}.  Here, the inspiral waveform
extends beyond the ISCO to a merger frequency, after which point the
amplitude evolves as $A(f) \propto f^{-2/3}$, and finally
incorporates a ringdown.  For $10 - 10 M_{\odot}$ waveforms, the ISCO is
at 220 Hz, but the phenomenological waveform continues up to 800 Hz.  We
will show that, by including the merger and ringdown information, the
timing and localization accuracies for these waveforms can be
improved dramatically. 

\section{Timing Accuracy}
\label{sec:timing}

The parameter estimation problem has been discussed in detail in many
articles.  Here, we provide a brief overview of the method in order to
fix notation, (for further details, see e.g. \cite{Maggiore:1900zz}).
We then proceed to use the framework to address the specific problem of
timing accuracy.  In later sections, we make use of the same framework
to obtain localization estimates and address systematic uncertainties.

\subsection{Parameter Estimation and the Fisher Matrix}
\label{ssec:param_est}

In order to decide whether there is a signal present in the data, we
calculate the likelihood ratio of a signal $h$ parametrized by some set
$\boldsymbol\mu$ of parameters%
\footnote{e.g.~for the coalescing binary signal introduced in section
\ref{ssec:cbc_waveform}, $\boldsymbol\mu =
(m_{1}, m_{2}, t_{o}, D, \theta, \phi, \psi, \iota, \phi_{o})$.}
 being present in the data $s$, relative to the
null hypothesis:
\begin{equation}\label{eq:likelihood}
  \Lambda(\boldsymbol\mu) = \frac{p(s|h(\boldsymbol\mu))}{p(s|0)} = 
  \frac{e^{- \langle s - h(\boldsymbol\mu) | s - h(\boldsymbol\mu) \rangle /2}}{ 
  e^{- \langle s| s \rangle /2}} \, , 
\end{equation}
where the inner product is defined as
\begin{equation}\label{eq:inner_product}
  \langle a | b \rangle = 4 \, \mathrm{Re} \int_{0}^{\infty} df
  \, \frac{\tilde{a}(f) \tilde{b}^{\star}(f)}{ S(f) } \, ,
\end{equation}
and $S(f)$ is the noise power spectrum of the detector.  The
formalism can be used both for the purposes of detection and parameter
estimation, as described in e.g. \cite{Jaynes03}.  For parameter
estimation, we are interested in the posterior probability distribution
of the parameters $\boldsymbol\mu$, given the data $s$.  The posterior
distribution for the parameters $\boldsymbol\mu$ can be obtained using
Bayes' theorem as:
\begin{equation}\label{eq:param_pdf}
  p(\boldsymbol\mu | s) = \frac{p(\boldsymbol\mu) \, p(s | \boldsymbol\mu)}{
    \int d\boldsymbol\mu \, p(\boldsymbol\mu) \, p(s | \boldsymbol\mu) }
   = \frac{ p(\boldsymbol\mu) \, \Lambda(\boldsymbol\mu) }{ 
    \int d\boldsymbol\mu \, p(\boldsymbol\mu) \, \Lambda(\boldsymbol\mu) }
\end{equation}
where $p(\boldsymbol\mu)$ is the prior distribution on the parameters
$\boldsymbol\mu$ and $\Lambda(\boldsymbol\mu)$ is the likelihood ratio
introduced above.  If we are only interested in a subset
$\boldsymbol\xi$ of the parameters $\boldsymbol\mu$ and not
$\boldsymbol\nu$, we simply marginalize over the ``nuisance'' parameters
$\boldsymbol\nu$ by integrating over them.

Let us now specialise to the case discussed in Section
\ref{sec:waveform} where the waveform is parametrized by two orthogonal
phases $h_{0}$ and $h_{\frac{\pi}{2}}$ satisfying
(\ref{eq:hf_relation}), with arbitrary amplitudes $A_{0}$ and
$A_{\frac{\pi}{2}}$ as in (\ref{eq:h_det}), and substitute this waveform
into the expression (\ref{eq:likelihood}) for the likelihood.  The
expression is simplified by noting that the two phases are necessarily
orthogonal,
\begin{eqnarray}\label{eq:orthogonal_phases}
  \langle h_{0}(\boldsymbol\xi) | h_{0}(\boldsymbol\xi) \rangle = 
  \langle h_{\frac{\pi}{2}}(\boldsymbol\xi) |
  h_{\frac{\pi}{2}}(\boldsymbol\xi) \rangle 
  \quad \mathrm{and} \quad 
  \langle h_{0}(\boldsymbol\xi) | h_{\frac{\pi}{2}}(\boldsymbol\xi) \rangle = 0 \, .
\end{eqnarray}
Finally, by either maximizing the likelihood with respect to $A_{0}$ and
$A_{\frac{\pi}{2}}$ or by marginalizing over them with a uniform prior,%
\footnote{In many cases, the sources of interest are approximately
uniformly distributed in volume.  This leads to a prior on the amplitude
of $A^{-4}$.  A uniform prior is chosen here for ease of calculation.
For observed signals, the amplitude will be large enough that the choice
of prior will not have a substantial effect on the parameter accuracy 
estimates derived later.} 
 we obtain
\begin{equation}\label{eq:log_like}
\ln \Lambda(\boldsymbol\xi) = 
   \frac{\langle s|h_{0}(\boldsymbol\xi) \rangle^{2} + 
   \langle s | h_{\frac{\pi}{2}}(\boldsymbol\xi)\rangle^{2} }{
   2 \langle  h_{0}(\boldsymbol\xi) | h_{0}(\boldsymbol\xi) \rangle } \, .
\end{equation}

For other parameters, it is not possible to handle the marginalization
of the likelihood so straightforwardly.  Therefore, we make use of the
Fisher Information Matrix to obtain expected parameter estimation
accuracies.  Briefly, assume that there is a signal in the data of the
form
\begin{equation}\label{eq:signal}
  s = A_{0} h_{0}\langle \boldsymbol\xi) + A_{\frac{\pi}{2}}
  h_{\frac{\pi}{2}}(\boldsymbol\xi) + n \, .
\end{equation}
Further, assume that the amplitude of the signal is sufficiently large
that the noise contribution $n$ can be neglected.  Then, expand the
likelihood in the neighbourhood of the true signal parameter in powers
of $d\boldsymbol\xi$.  Setting $\langle h_{0}(\boldsymbol\xi) |
h_{0}(\boldsymbol\xi)\rangle  = 1$ we obtain
\begin{equation}\label{eq:quad_like}
    \ln \Lambda(d \boldsymbol\xi) \approx 
    \frac{\rho^{2}}{2} \left[ 1  -  g_{ab} d\xi^{a} d\xi^{b} \right]
\end{equation}
where $\rho^{2} = (A_{0}^{2} + A_{\frac{\pi}{2}}^{2})$ and 
\begin{equation}\label{eq:metric} 
  g_{ab} = \langle  \partial_{a} h_{0}| \partial_{b} h_{0}\rangle  
    - \langle  h_{0} | \partial_{a} h_{0}\rangle  \langle  h_{0} | \partial_{b} h_{0}\rangle   
    - \langle  h_{\frac{\pi}{2}} | \partial_{a} h_{0}\rangle  
    \langle  h_{\frac{\pi}{2}} | \partial_{b} h_{0}\rangle  \, ,
\end{equation}
is a positive definite matrix.  At quadratic order, the likelihood
function is approximated as a multi-variate Gaussian around the peak
$d\boldsymbol\xi = 0$. The Fisher matrix $g_{ab}$ then provides an estimate
of the accuracy with which the parameters $\boldsymbol\xi$ can, in
principle, be determined.  

\subsection{Timing}
\label{ssec:timing}

It is rather straightforward to utilize the formalism introduced above
to investigate the effects of a timing error.  In particular, restrict
attention to the case where the parameter space $\boldsymbol\xi$ is the
one dimensional time parameter.  In this case, a shift of the time $t_{o}$
corresponds to (frequency dependent) phase shift in the waveform and has
no effect on the amplitude of the waveform, specifically
\begin{equation}
  \tilde{h}_{0}(f; t_{o} + dt ) = 
  e^{2 \pi i f dt} \tilde{h}_{0}(f; t_{o}) \, .
\end{equation}

We can obtain a posterior distribution for the timing error, making use
of the formalism introduced above.  Generally, the time will not be
known apriori, certainly not to millisecond accuracy, so it is natural
to take a uniform prior $p(dt) = \mathrm{const}$.  Then, the posterior
distribution for the time offset is 
\begin{equation}\label{eq:log_like_dt}
p(d t | s) \propto \exp \left\{ 
   \frac{\rho^{2}}{2} \left[ \langle h_{0}|h_{0}(d t)\rangle^{2} + 
  \langle  h_{\frac{\pi}{2}} | h_{0}(d t)\rangle ^{2} \right]
\right\}\, .
\end{equation}
At quadratic order, this gives%
\footnote{This expression is somewhat different that what is obtained by
directly expanding Eq.~(\ref{eq:log_like_dt}).  However, by expanding $\langle 
h(dt) | h(dt) \rangle  = 1$ in powers of $dt$, it is easy to show
that the two expressions are equivalent.}%
\begin{eqnarray}
  \langle h_{0}|h_{0}(d t)\rangle ^{2} &\approx& 1 - dt^{2} \left[ 
  \langle \partial_{t} h_{0} | \partial_{t} h_{0}\rangle  
  - \langle  h_{0} | \partial_{t} h_{0} \rangle ^{2} \right] \label{eq:alpha}\\
  \langle  h_{\frac{\pi}{2}} | h_{0}(d t)\rangle ^{2}
  &\approx& dt^{2} \langle h_{\frac{\pi}{2}} | \partial_t h_{0} \rangle ^{2} \, .
\label{eq:beta}
\end{eqnarray}
Then, since taking the derivative of the waveform with respect to $t_{o}$
amounts to multiplication by $2\pi i f$, 
\begin{equation}
  \langle h_{0}|h_{0}(d t)\rangle ^{2} 
  \approx 1 - 4 \pi^{2} dt^{2} \langle f h_{0} | f h_{0} \rangle 
  = 1 - 4 \pi^{2} (d t)^{2} \overline{f^{2}} \, .
\end{equation}
Here, we have introduced $\overline{f^{n}}$ to describe the frequency
moments of the signal as 
\begin{equation}\label{eq:fn_mean}
  \overline{f^{n}} := 4 \int_{0}^{\infty} df \frac{ | \tilde{h}(f)
|^{2}}{S(f)} f^{n} \, .
\end{equation}
Similarly (\ref{eq:beta}) is approximated as
\begin{equation}
  \langle  h_{\frac{\pi}{2}} | h_{0}(d t)\rangle 
  \approx 2 \pi \overline{f} dt \, ,
\end{equation}
where $\overline{f}$ is the mean frequency defined via
(\ref{eq:fn_mean}).  Thus, at quadratic order, the timing distribution
is
\begin{equation}\label{eq:timing_dist}
  p(dt | s) \propto \exp 
  \left[ - 2 \rho^{2}\pi^{2} \sigma_{f}^{2} dt^{2} \right]   
  \quad \mathrm{where} \quad
  \sigma_{f}^{2} = \overline{f^{2}} - \overline{f}^{2}
\end{equation}
is the \textit{effective bandwidth} of the signal.  In this case, the
timing estimator is unbiased and has a width given by
\begin{equation}\label{eq:sigma_t}
  \sigma_{t} = \frac{1}{2 \pi \rho \sigma_{f}} \, .
\end{equation}
This is a simple result which encapsulates the expected timing accuracy
for a given source.  It is inversely proportional to both the signal to
noise ratio (SNR) $\rho$ and effective bandwidth $\sigma_f$ of the source.

\subsection{Example: Binary Coalescence}
\label{ssec:bns_timing}

We can apply the results obtained above to the binary coalescence
waveforms of section \ref{ssec:cbc_waveform}.  As well as providing a
concrete example, it will allow us to investigate the accuracy of the
quadratic approximation used to derive Eq.~(\ref{eq:sigma_t}) above.

We begin by considering a $1.4 - 1.4 M_{\odot}$ BNS.  The values of the
mean frequency $\bar{f}$, effective bandwidth $\sigma_{f}$ and timing
accuracy $\sigma_{t}$ are given in Table \ref{tab:bns_time}.  Interestingly,
all of the detectors have similar mean frequencies and bandwidths with
the broad design noise curve of Virgo leading to the largest effective
bandwidth.  Note that although the frequency bandwidth of Advanced LIGO
is not significantly greater than initial LIGO, a signal at the same
distance would appear with approximately twelve times the SNR, and
therefore the timing accuracy would be fifteen times better.  A detailed
study of the timing accuracy for various different Advanced LIGO
configurations has been investigated in \cite{King:2008}.

\begin{table}[t]
\center
\begin{tabular}{| c | c | c | c | c | c | c | }
\hline \hline
Detector  & $\overline{f}$ (Hz) & $\sigma_{f}$ (Hz) & 
  \multicolumn{2}{c|}{ $\sigma_{t}$(ms)} & \multicolumn{2}{c|}{ Timing
accuracy 
(ms)}\\ 
 & & & $\rho = 7$ & $\rho = 10$ & $\rho = 7$ & $\rho = 10$ \\
\hline
\hline
Initial LIGO  & 150 & 100 & 0.23 & 0.17 & 0.27 & 0.18 \\
Initial Virgo & 140 & 140 & 0.16 & 0.11 & 0.19 & 0.12 \\
Advanced LIGO & 110 & 120 & 0.19 & 0.13 & 0.21 & 0.13 \\
\hline
\hline
\end{tabular}
\caption{Timing accuracy for binary neutron stars in different
detectors.  The table gives the mean frequency and effective bandwidth
of the signal.  Then, the timing accuracies based on the quadratic
approximation (\ref{eq:sigma_t}) and the exact result are given for
SNR 7 and 10.}
\label{tab:bns_time}
\end{table}

\begin{figure}[ht]
\centering
\includegraphics[width=.45\textwidth]{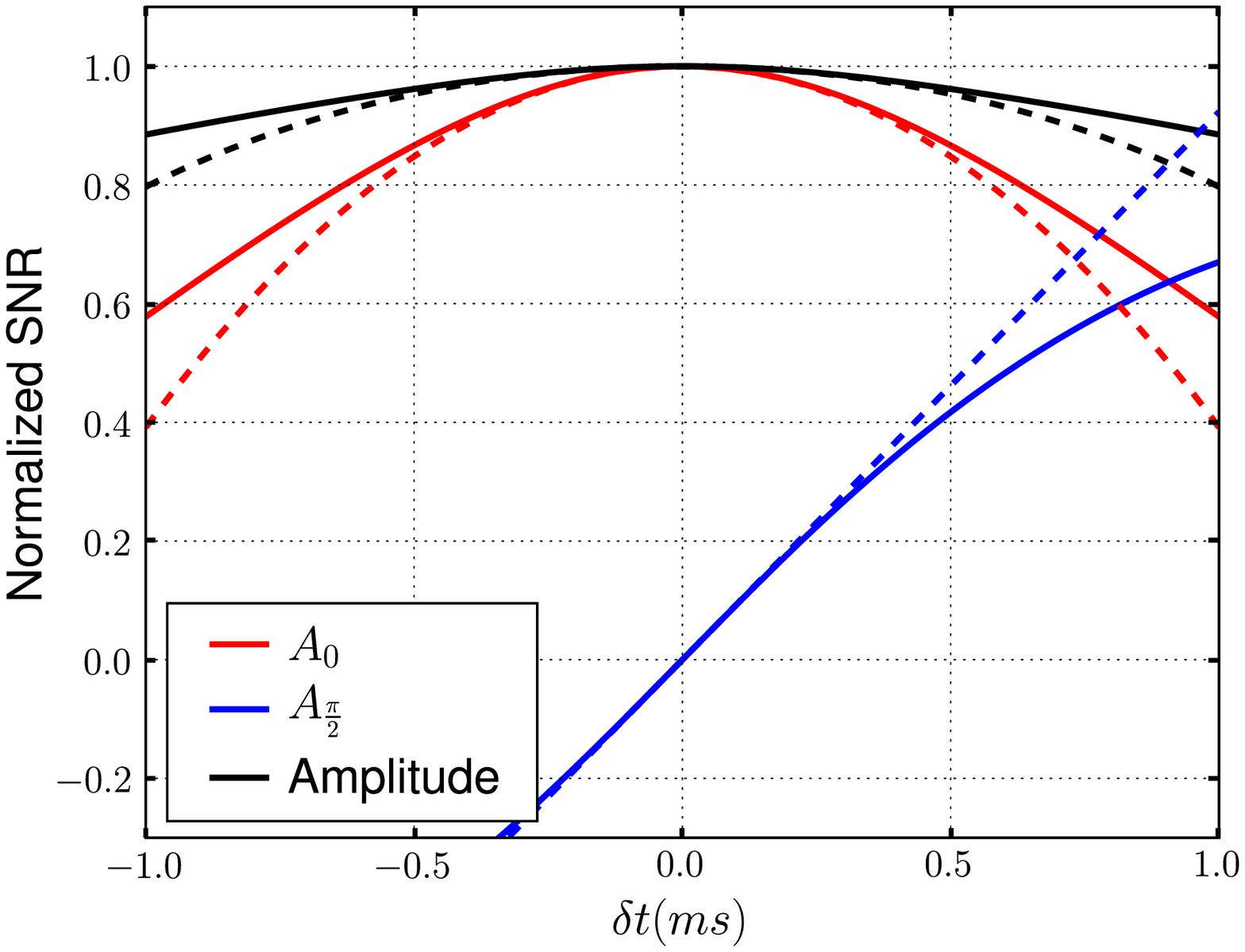}
\includegraphics[width=.45\textwidth]{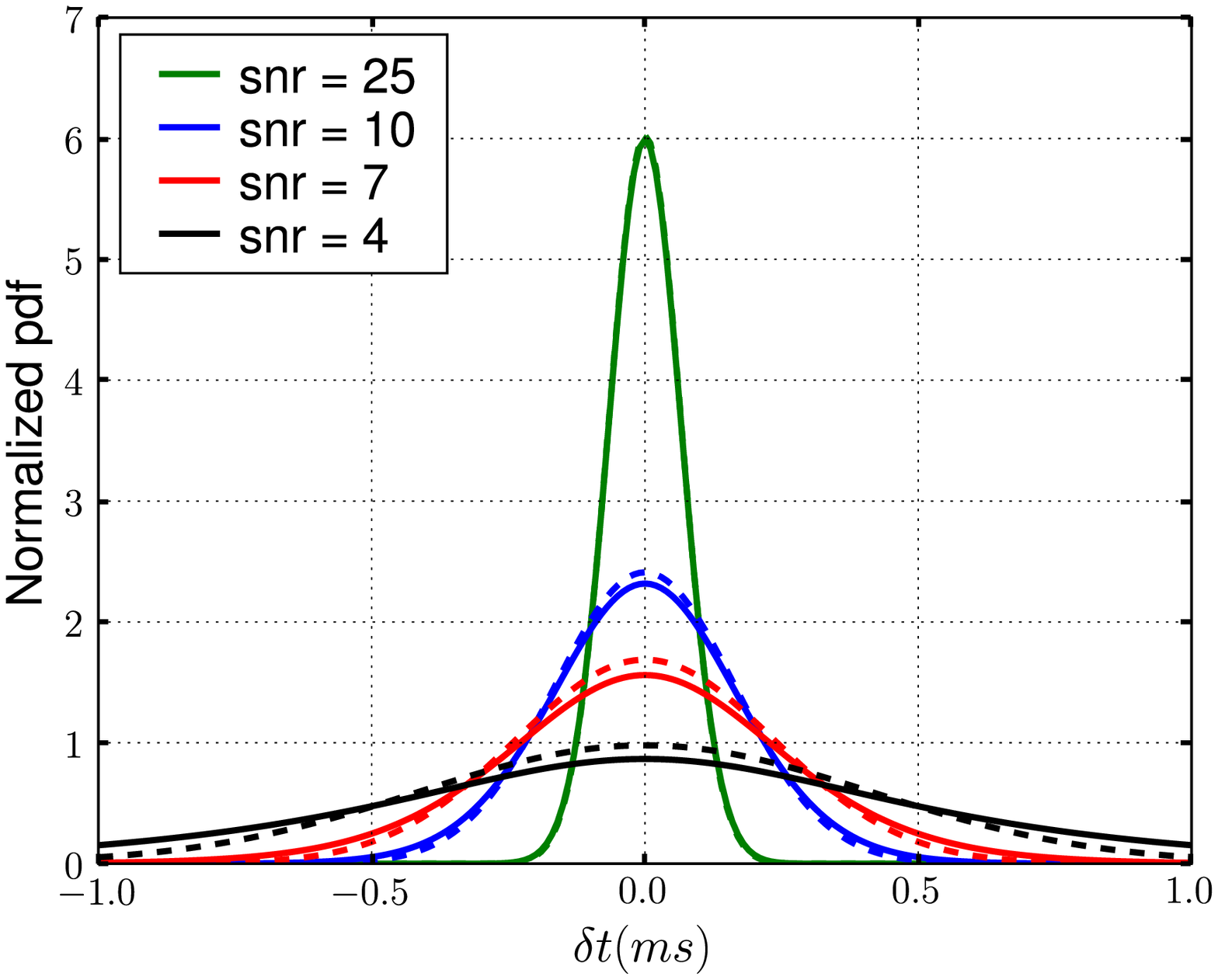}
\caption{\textbf{Left:} The normalized SNR ratio against
time for initial LIGO.  The SNR for the two phases of the template are
plotted, as a function of time, when the waveform has zero phase.  The
solid lines are the exact results, dotted lines show the SNR
approximated using the quadratic approximation discussed in this
section. \textbf{Right:} The timing distribution for a BNS system at a
given SNR in initial LIGO.  The dotted lines are the quadratic
approximation, while the solid curves are the exact expression. At low
SNR, the quadratic approximation underestimates the timing uncertainty.}
\label{fig:snr_time}
\end{figure}

In Figure \ref{fig:snr_time}, we investigate the accuracy of the
quadratic approximation used in obtaining (\ref{eq:sigma_t}).  The
figure shows the SNR as a function of time for initial LIGO, exactly
calculated and using the approximation introduced above.  In both cases,
the masses of the simulated signal and template waveform agree.  The
quadratic approximation is only good to about $0.5$ ms after which it
significantly underestimates the recovered SNR.  This leads to an
underestimation of the timing uncertainty, as shown in the right
hand plot.  The approximated distributions are more sharply peaked than
the exact ones; the timing uncertainty is underestimated by about
$20\%$ at a signal SNR of 7, and $15\%$ at SNR of 10.  It is only at SNR
of 25 or more that the quadratic approximation introduces negligible
error.  

\begin{table}[t]
\center
\begin{tabular}{| c | c | c | c | c | c | c | c |}
\hline \hline
 & \multicolumn{3}{c|}{Waveform to ISCO} 
 & \multicolumn{4}{c|}{Full waveform} \\ 
  & $\overline{f}$ (Hz) & $\sigma_{f}$ (Hz) & $\sigma_{t}$(ms)  
  & $\overline{f}$ (Hz) & $\sigma_{f}$ (Hz) & $\sigma_{t}$(ms) & 
  $\sigma_{t}$(ms) \\ 
Detector  
  &  &  & $\rho = 10$ & & & $\rho = 10$ & $\rho_{I} = 10$ \\ 
\hline
\hline
Initial LIGO  & 120 & 40 & 0.37 & 150 & 100 & 0.16 & 0.14 \\
Initial Virgo &  90 & 50 & 0.34 & 140 & 140 & 0.12 & 0.10 \\
Advanced LIGO &  75 & 50 & 0.34 & 120 & 130 & 0.11 & 0.10  \\
\hline
\hline
\end{tabular}
\caption{Timing accuracy based on quadratic approximation for $10-10
M_{\odot}$ black hole binary.  The results are given for the waveform
truncated at the ISCO frequency and for the full waveform based on the
phenomenological waveforms of \cite{Ajith:2007qp}.  We give the timing
accuracy for signals at SNR 10.  The final column shows the timing
accuracy for a waveform which accumulates an SNR of 10 to ISCO, if found
using the full waveform template.}  
\label{tab:bbh_time}
\end{table}

Finally, consider a $10-10 M_{\odot}$ BBH system.  Table
\ref{tab:bbh_time} gives the effective bandwidth and timing accuracy for
both post--Newtonian waveforms truncated at ISCO and full,
phenomenological waveforms.  Interestingly, even though only 15 to 20\%
additional SNR is accumulated after ISCO, the timing accuracy improves 
by as much as a factor of three due to the additional high frequency
content of the waveform.

\section{Sky Localization from Triangulation}
\label{sec:location}

Given a timing uncertainty in each of a network of detectors, this can
be translated to a localization accuracy for the network.  In the
previous section, we have derived a simple expression for the timing
accuracy for an elliptical waveform as $\sigma_{t} \approx (2\pi \rho
\sigma_{f})^{-1}$, where $\rho$ is the observed SNR and $\sigma_{f}$ is
the effective bandwidth of the signal in the detector.  However, much of
what follows makes use only of the timing accuracy, without reference to
its derivation and would be applicable to other waveform families. 

To obtain the timing results, we have assumed that the amplitudes of the
two phases of the waveform in each detector are independent.  While this
is valid for a single detector, for more than two detectors, these
amplitudes are not independent since the gravitational wave has only two
polarizations.  This can be seen in detail for coalescing binaries using
a simple counting argument.  For three detectors we make 9 measurements
(two amplitudes and a time in each), but these are dependent on only 7
parameters $(D, \theta, \phi, \psi, \iota, \phi_{o}, t_{o})$.  However,
it is reasonable to assume that any correlations between observed
detector amplitudes will only serve to improve the accuracies derived
below.

\subsection{Two site network}
\label{sec:2det}

A two site network will give single measurement of timing differential
and will therefore provide only partial localization of the signal.
Suppose that the source is located at position $\mathbf{R}$ on the unit
sphere, and consider two detectors separated by a distance (expressed in
light seconds) of $\mathbf{D}$.  Then, the difference in the time of
arrival of the signal between the two sites is
\begin{equation}\label{eq:rx}
  (T_{1} - T_{2}) = \mathbf{D} \cdot {\mathbf{R}}  
\end{equation}
If the two detectors have timing accuracies $\sigma_{1}$ and
$\sigma_{2}$, the distribution of the observed times $t_{1}$ and $t_{2}$
is 
\begin{equation}\label{eq:2det_pdf}
  p(t_{1}, t_{2} | s) \propto  p(t_{1}, t_{2})
  \exp \left[
  - \frac{(t_{1} - T_{1})^{2}}{2 \sigma_{1}^{2}} 
  - \frac{(t_{2} - T_{2})^{2}}{2 \sigma_{2}^{2}} 
    \right] \, .
\end{equation}
Localization will depend only upon the time delay $(t_{1} - t_{2})$, so
we re-express (\ref{eq:2det_pdf}) in terms of a fiducial arrival time
and the reconstructed location $\mathbf{r}$.  Marginalizing over the
arrival time, with a uniform prior, gives 
\begin{equation} p(\mathbf{r} | \mathbf{R}) \propto p(\mathbf{r}) \exp
\left[- \frac{ \left( \mathbf{D} \cdot \left( \mathbf{r} -
\mathbf{R}\right) \right)^{2}}{ 2 (\sigma_{1}^{2} + \sigma_{2}^{2}) }
\right] \, .  \end{equation}
As expected, from the timing observation in two detectors, it is only
possible to restrict the location of the source in the direction
parallel to the separation $\mathbf{D}$ between the detectors.  The
localization ability is improved by better timing accuracy in the
detectors, and also by an increased baseline between detectors.  

When localizing a source, we would like to provide the smallest region
of the sky which contains the source, with a given confidence.  This
requires the choice of a prior distribution for the sky location
$\mathbf{r}$.  In the absence of directional information,%
\footnote{In some cases, it might be reasonable to change this
assumption.  For example, in many cases it is reasonable to restrict the
prior on $\mathbf{r}$ to be localized to nearby galaxies
\cite{LIGOS3S4Galaxies}.  Alternatively, the detectors' directional
sensitivities make is more likely that the signal came from certain sky
locations.}%
 it is natural to choose the prior on $\mathbf{r}$ to be uniformly
distributed on the unit sphere, giving a uniform prior distribution of
$\mathbf{D} \cdot \mathbf{r} \in [-1, 1]$.  Then, for a 90\% confidence
region, we obtain
\begin{equation}
  \frac{\mathrm{Area}(90\%)} {4 \pi} \approx 
  \frac{3.3 \sqrt{\sigma_{1}^{2} + \sigma_{2}^{2}}}{D} \, .
\end{equation}
The area is independent of the location of the signal.  For the LIGO
detectors $D = 10$ ms so that a timing accuracy of $0.25$ ms in each
detector limits the signal to about 12\% of the sky.  For LIGO and
Virgo, the light travel time is significantly larger at $27$ ms and
consequently, with the same timing accuracy, the signal can be localized
to about 4\% of the sky.  

\subsection{Three site network}
\label{sec:3det}

The three detector result can be obtained in a similar manner.  As
before, we re-express the observed detector arrival times in terms of
the reconstructed sky location $\mathbf{r}$ and fiducial arrival time
$t_{0}$.  Marginalizing over the arrival time gives 
\begin{equation}\label{eq:3site_post}
  p(\mathbf{r} | \mathbf{R} ) \propto p(\mathbf{r}) 
  \exp \left[ - \frac{1}{2} (\mathbf{r} - \mathbf{R})^{T} \mathbf{M} 
  (\mathbf{r} - \mathbf{R}) \right] \, ,
\end{equation}
where the matrix $\mathbf{M}$, describing the localization accuracy, is
given by
\begin{equation}
  \textbf{M} = \frac{\mathbf{D}_{12} \mathbf{D}_{12}^{T}}{\sigma_{12}^{2}} 
    + \frac{\mathbf{D}_{23} \mathbf{D}_{23}^{T}}{\sigma_{23}^{2}} 
    + \frac{\mathbf{D}_{31} \mathbf{D}_{31}^{T}}{\sigma_{31}^{2}} \, .
\end{equation}
Thus $\mathbf{M}$ has a contribution from each pair of detectors which
depends upon the detector separation $\mathbf{D}_{ij}$ and the pairwise
timing uncertainty
\begin{equation}
  \sigma_{ij}^{2} = \sigma_{i}^{2} + \sigma_{j}^{2} 
    + \frac{\sigma_{i}^{2} \sigma_{j}^{2}}{\sigma_{k}^{2}} 
\end{equation}
where $k \neq i,j$.  The timing uncertainty from a given pair of
detectors is dependent upon the timing accuracy $\sigma_{k}$ in the
third detector.  Initially, this may seem surprising, but arises quite
naturally due to the single marginalization over the fiducial arrival
time.  Finally, we note that the two detector result (\ref{eq:2det_pdf})
can be reproduced by taking $\sigma_{3} \rightarrow \infty$. 

Since the $\mathbf{D}_{ij}$ are coplanar, $\mathbf{M}$ will have a zero
eigenvalue and hence a degenerate direction $\hat{e}_{z}$ normal to this
plane.  Thus, the three detector network can only restrict the sky
location projected onto the plane formed by the detectors.  In addition,
since $\mathbf{M}$ is independent of the sky location --- it depends
solely on the location and timing accuracies of the individual detectors
--- the localization ability \textit{within the plane of the detectors}
is independent of the location in that plane.  We can complete the
co-ordinate system by introducing co-ordinates $\hat{e}_{x}$ and
$\hat{e}_{y}$ in eigen-directions of $\mathbf{M}$.  In this basis, the sky
localization distribution is
\begin{equation}\label{eq:prob_xy}
  p(\mathbf{r} | \mathbf{R} ) \propto p(\mathbf{r})
  \exp \left[ - \frac{1}{2} \left( 
  \frac{\left( x - X \right)^{2}}{\sigma_{x}^{2}} +
  \frac{\left( y - Y \right)^{2}}{\sigma_{y}^{2}} 
  \right) \right] \, ,
\end{equation}
where $(X,Y)$ are the co-ordinates of the source, projected onto the
plane of the detector and $(x,y)$ describe the recovered location.  

As before, we will use this distribution to obtain confidence regions on
the sky.  The regions will depend upon the prior distribution on
$\mathbf{r}$.  Although a uniform distribution on the unit sphere does
not lead to a uniform distribution on the $x-y$ plane, in most cases the
source localization is sufficiently accurate to treat $p(x,y)$ as
constant over this small region; we will make this approximation.  In
projecting the result back to the sky, we obtain two mirror sky
locations ($z \rightarrow -z$) and an additional factor of $(\cos
\theta)^{-1}$, where $\theta$ is the angle between the line of sight and
the normal to the plane of the detectors.%
\footnote{Of course this breaks down when the source approaches the
plane of the detectors, namely $\theta \sim \pi/2$.  Close to the plane
of the detectors, we can approximate the sky localization by considering
the extreme case where the source is in the plane of the detectors, and
specifically at $x = 1, y = 0$.  The uncertainty in the y-direction will
be proportional to $\sigma_{y}$ but in the z-direction, we obtain
$\sigma_{z} \propto \sqrt{\sigma_{x}}$.}
If we assume that the mirror degeneracy can be broken with amplitude
consistency tests, the source can be localized with probability $p$ to
an area:
\begin{equation}
  \mathrm{Area}(p) \approx 2 \pi \sigma_{x} \sigma_{y} \left[ - \ln (1 -
p) \right]/ \cos{\theta} \, .
\end{equation}
If the reflection degeneracy cannot be broken then the area is doubled,
apart from close to the plane of the detectors in which case the error
boxes from the mirror locations will overlap.  The best case scenario
occurs when the signal is directly overhead the plane of the detectors.
The median occurs when $\cos \theta = 1/2$ and gives a factor of two
increase in localization area.  

To provide a concrete example of expected localization abilities, we
consider the LIGO-Virgo network of detectors.  The light travel time
between the LIGO sites is $10$ ms, while between Virgo and the two LIGO
sites is around $27$ ms for both.   Since the two LIGO detectors have
similar sensitivities and a very similar orientation, a large fraction
of sources will have similar timing uncertainties in the two detectors.
Therefore, we simplify by taking the timing accuracy $\sigma_{l}$ of
the two LIGO detectors to be identical, but allow the Virgo timing
$\sigma_{v}$ to differ.  Then, the eigen-directions of the matrix
$\mathbf{M}$ are roughly aligned with the line connecting the LIGO sites
$\hat{e}_{x}$ and the line connecting its midpoint to Virgo
$\hat{e}_{y}$ and the localization accuracies in these directions are given
by
\begin{equation}\label{eq:sigma_xy}
  \sigma_{x} \approx \frac{\sigma_{l}}{7 \mathrm{ms}} 
  \quad \mathrm{and} \quad 
  \sigma_{y} \approx \frac{\sqrt{(2 \sigma_{v}^{2} +
\sigma_{l}^{2})/3}}{22 \mathrm{ms}} \, .
\end{equation}

For a 90\% confident localization, assuming reflection degeneracy can be
broken
\begin{equation}\label{eq:lv_loc}
  \mathrm{Area}(90\%,\mathrm{best}) \approx 20 \deg^{2} 
    \left(\frac{\sigma_{l}}{0.25 \mathrm{ms}} \right)
    \left(\frac{\sqrt{(2 \sigma_{v}^{2} +
\sigma_{l}^{2})/3}}{0.25 \mathrm{ms}} \right)
\end{equation}
The localization error box is double the size for the median location,
$\cos \theta = 1/2$, and will contain two equal size, disconnected
pieces if the reflection degeneracy cannot be broken.  For the worst
case scenario, where the source is in the plane of the detectors, we
obtain an error box of over $100 \deg^{2}$ for a $0.25$ ms timing
accuracy. For  reference, a $1 \deg^{2}$ localization for best and
median sources requires timing accuracies of $0.06$ ms, $0.04$ ms in all
detectors.

\subsection{Example: Binary Coalescence}
\label{sec:cbc_loc}

Based on the results of recent searches \cite{Abbott:2009tt}, we
will take an SNR of 7 in each of the LIGO and Virgo detectors to be the
approximate amplitude where a binary coalescence signal would stand
above the noise background.  For BNS signals of this amplitude Table
\ref{tab:bns_time} gives a timing accuracy of $0.27$ ms for the LIGO
detectors and $0.19$ ms for Virgo.  This gives at best case localization
of $20 \deg^{2}$.  A signal would require an SNR of around $25$ and a
well located source to reduce the $90\%$ localization ellipse to $1
\deg^2$ --- certainly a possibility for the louder sources in the
advanced detector era.  The localization accuracy for $10 - 10
M_{\odot}$ BBH waveforms is comparable to BNS, namely $20 \deg^{2}$ for
optimally located signals at single detector SNR of 7.  Interestingly,
the inclusion of the merger and ringdown portions of the signal provide
an order of magnitude improvement in the localization accuracy.  This
improvement is consistent with what was observed in a detailed study of
parameter estimation for BBH \cite{Ajith:2009fz}.

\begin{figure}[ht]
\centering
\includegraphics[width=.60\textwidth]{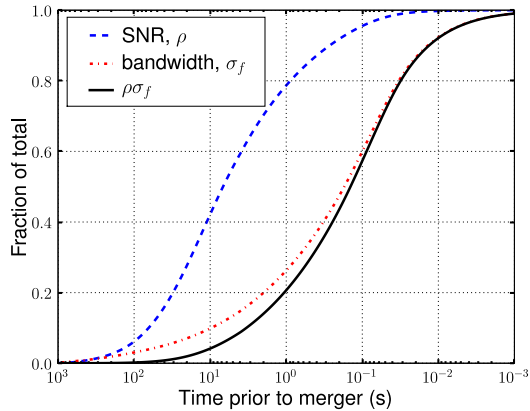}
\caption{ The growth of SNR $\rho$ and signal bandwidth $\sigma_{f}$ as
a function of time prior to merger.  The timing uncertainty$\sigma_{t}
\propto (\rho \sigma_{f})^{-1}$ decreases as the binary nears merger.
Although a large fraction of the SNR is accumulated well before merger,
the bandwidth increases significantly right before merger.  Hence, the
timing uncertainty is only decreased in the last second or so prior to
merger.}
\label{fig:adligo_pre_merger}
\end{figure}

The sensitive band of advanced detectors is expected to begin at around
10 Hz, and it it will take a BNS system over a thousand seconds to
evolve from 10 Hz to coalescence.  Thus, it is interesting to consider
the possibility of localizing the source \textit{prior} to detection to
allow for early pointing of electromagnetic telescopes.  The SNR of the
signal $\rho(t)$ will accumulate during the coalescence.  Similarly, the
frequency will evolve.  During the inspiral phase, the frequency of the
orbit evolves to leading order as $f(t) \propto (t_{o} - t)^{-3/8}$
where $t_{o}$ is taken to as time of coalescence.  Thus, making use of
(\ref{eq:fn_mean}) and (\ref{eq:timing_dist}) we can calculate the
accumulated bandwidth of the signal as a function of time,
$\sigma_{f}(t)$.  Using (\ref{eq:sigma_t}), we can investigate how the
timing accuracy $\sigma_{t}$ evolves during the coalescence.

Figure \ref{fig:adligo_pre_merger} shows the evolution of the
accumulated SNR, frequency bandwidth and timing accuracy over the course
of the binary inspiral.  While a large fraction of SNR has
accumulated 10 seconds before merger, the bandwidth, and consequently
the timing accuracy, is largely accrued in the last second prior to
merger.  Unfortunately, this would seem to make advanced localization of
BNS systems unlikely, even with advanced detectors.  However,
electromagnetic follow-up is still worthwhile as signals associated with
coalescing binaries are expected to be emitted after coalescence
\cite{NakarReview:2007}, or be delayed by dispersion through the
interstellar medium \cite{Lorimer:2007qn} and therefore arrive after the
gravitational wave signal.

\section{Systematic Uncertainties}
\label{sec:systematics}

In a gravitational wave search, there are numerous systematic errors
associated to the waveform which affect the localization accuracy.  In
this section, we consider systematic uncertainties due to waveform and
calibration errors.  These issues have been considered in the context of
detectability and parameter estimation accuracy in
\cite{Lindblom:2008cm, Lindblom:2009ux, Lindblom:2009un}.  Errors in the
template waveform might arise due to the breakdown of analytic
approximations, or numerical inaccuracies in simulating the waveform.
We also consider the effect of calibration errors on localization and
show that they can be handled in a very similar manner to waveform
errors.  However, calibration errors are independent at the different
detectors, whereas the waveform errors are not.

\subsection{Waveform errors}
\label{sec:waveform_error}

Let us generalize the analysis introduced in section \ref{sec:timing}
by allowing for an error in the waveforms used for filtering.  Following
\cite{Allen:1996}, we write:
\begin{equation}\label{eq:waveform_errors}
  \tilde{h}_{0}(f; t_{o}) = A(f) (1 + \delta a) 
    \exp[2 \pi i f t_{o} +  i \Phi(f) + i \delta \phi(f)] \, ,
\end{equation}
where $\delta a$ and $\delta \phi$ characterize the amplitude and phase
errors in the waveform respectively; we continue to assume that
$\tilde{h}_{\frac{\pi}{2}}(f; t_{o}) = i \tilde{h}_{0}(f; t_{o})$. Since
$\delta a$ and $\delta \phi$ are unknown functions of frequency, they
cannot be treated with the standard Fisher matrix technology.  However
we can expand the likelihood in powers of $\delta a$ and $\delta \phi$
to second order to obtain
\begin{eqnarray}\label{eq:amp_phase_like}
  \ln \Lambda(\delta a, \delta \phi, \delta t) \approx \frac{\rho^{2}}{2}
  [ 1 
  &-& \langle h_{0} (2\pi f \delta t + \delta \phi) | 
     h_{0} (2\pi f \delta t + \delta \phi)\rangle  \nonumber \\
  &+& \langle h_{0} (2\pi f \delta t + \delta \phi) | h_{0}\rangle^{2} 
     \nonumber \\
  &-& \langle h_{0} (\delta a) | h_{0} (\delta a) \rangle
  + \langle h_{0} (\delta a) | h_{0} \rangle^{2} ) \, .
\end{eqnarray}
Interestingly, there are no cross terms between the amplitude and phase
errors.  Therefore, although the amplitude errors will affect the
likelihood they do not have an effect (at leading order) on the timing.

Given a phasing error $\delta \phi$, it is straightforward to
differentiate (\ref{eq:amp_phase_like}) to obtain the timing offset
$\widehat{\delta t}$ which maximizes the likelihood as
\begin{equation}\label{eq:timing_systematic}
  \widehat{\delta t} = \frac{1}{\sigma_{f}^{2}} 
  \left( 4 \int_{0}^{\infty} df 
  \frac{ | \tilde{h}(f) |^{2}}{S(f)} (\bar{f} - f)
  \left[\frac{\delta \phi(f)}{2 \pi}\right] \right) \, .
\end{equation}
In general the phasing error $\delta \phi$ is not known explicitly,
instead bounds on the maximum error $\delta \phi_{\mathrm{max}}$ are
usually provided.  In many cases only a maximum phase error is provided,
independent of frequency, then the timing offset is bounded by
\begin{equation}\label{eq:time_off}
  |\widehat{\delta t}| \le \frac{\delta \phi_{\mathrm{max}}}{
  2 \pi \sigma_{f}^{2}} 
  \left( 4 \int_{0}^{\infty} df \frac{ | \tilde{h}(f) |^{2}}{S(f)} 
  |\bar{f} - f| \right) \, ,
\end{equation}
and the bound is only obtained if the phase error is maximal at all
frequencies, and changes sign at the mean frequency $\bar{f}$.  Finally,
we note that the integrand in (\ref{eq:time_off}) gives the mean
absolute deviation of the frequency.  Since this is always less than or
equal to the standard deviation, the timing offset can be bounded by 
\begin{equation}\label{eq:loose_timing}
  |\widehat{\delta t}| \le \frac{1}{\sigma_{f}}
  \left[\frac{\delta \phi_{\mathrm{max}}}{2 \pi}\right] \, . 
\end{equation}
For most realistic phase errors, the uncertainty obtained from
(\ref{eq:timing_systematic}) will be significantly smaller than
(\ref{eq:loose_timing}).

The systematic error ($\widehat{\delta t}$ from
Eq.~(\ref{eq:loose_timing})) and statistical error ($\sigma_{t}$ from
Eq.~(\ref{eq:sigma_t})) are directly comparable.  Both are inversely
proportional to the frequency bandwidth of the signal.  The statistical
uncertainty is also inversely proportional to the amplitude (or SNR) of
the signal, while the systematic is independent of amplitude.
Thus, for a phasing error of $5^{\circ}$, the systematic offset is
guaranteed to be smaller than the statistical fluctuations for waveforms
with an SNR less than 12.  In reality, the bound in
(\ref{eq:loose_timing}) is rather loose and therefore a $5^{\circ}$
phasing error is unlikely to dominate the statistical timing uncertainty
at an SNR less than 20.  

The same waveform family will be used to search the data from all
instruments.  Therefore, the waveform error $\delta \phi$ will be the
same at all sites.  Two detectors will record a different timing offset
only if their power spectra (and consequently the mean frequency and
effective bandwidth) differ.  This immediately argues that between LIGO
sites with comparable sensitivities, the timing offset due to a waveform
error will be negligible.  The timing uncertainty between LIGO and
Virgo will depend upon the details of the waveform.  Given a specific
waveform and phase error, one can evaluate (\ref{eq:timing_systematic})
to obtain the results.

\subsection{Calibration Errors}
\label{sec:calibration_error}

Errors in the calibration of the detectors will also affect the timing
accuracy.  We have denoted the detector output or strain as s(t).  This
is obtained by calibrating the detector output $v(t)$ using a response
function $R(f)$.  This response function is then used to obtain the
calibrated data:\cite{Dietz:2006}%
\footnote{In practice, the process is generally performed in the time
domain to directly produce s(t).  However, that will not affect the
discussion below as the same systematic uncertainties in calibrating the
data still arise.}  
\begin{equation}\label{eq:response}
  \tilde{s}(f) = R(f) \tilde{v}(f) \, . 
\end{equation}
In a real detector, the response function is time dependent, however we
assume that the response function is (approximately) constant over the
duration of the signal.  In practice, the response function $R_{c}(f)$
which is calculated will not agree identically with the true response of
the instrument.  Using the measured response function, the instrumental
output is calibrated to obtain the data stream:

\begin{equation}\label{eq:miscalibrated_signal}
  \tilde{s}_{c}(f) = R_{c}(f) \tilde{v}(f) 
  = \frac{R_{c}(f)}{R(f)} \tilde{s}(f) \, .
\end{equation}
The errors in calibration will affect signals in the data, as well as
the noise and consequently the noise power spectrum, $S(f)$.

To calculate the effect of the calibration errors, we need to evaluate
the likelihood (\ref{eq:log_like}) obtained when using the incorrect
response function.  This requires the calculation of both the inner
product between signal and template, as well as the template
normalization, with the incorrect response function.  We will use the
notation $\langle \cdot | \cdot \rangle_{c}$ to denote the existence of
calibration errors in the power spectrum used in computing the inner
product.  

We begin by calculating the effect of calibration errors on the inner
product between signal and template:
\begin{eqnarray}
  \langle s_{c} | h \rangle_{c} 
  &=& 4  \, \mathrm{Re} \int df \, 
  \frac{\left(\frac{R_{c}(f)}{R(f)}\right) \tilde{s}(f) 
  \tilde{h}^{\star}(f)}
  { \left|\frac{R_{c}(f)}{R(f)} \right|^2 S(f)} \nonumber \\
  &=& 4 \, \mathrm{Re} \int df \, 
  \frac{\tilde{s}(f) \left(\frac{R(f)}{R_{c}(f)} \,
  \tilde{h}(f) \right)^{\star}}{S(f)} 
  = \langle s | h_{c} \rangle \, .
\end{eqnarray}
where we have introduced an effective waveform error due to calibration
as:
\begin{equation}
  h_{c}(f) = \left(\frac{R(f)}{R_{c}(f)}\right) \tilde{h}(f) \, .
\end{equation}
Similarly, it is straightforward to show that calibration errors in the
calculation of the template norm can be expressed as 
\begin{equation}
  \langle h | h \rangle_{c} = \langle h_{c} | h_{c} \rangle \, .
\end{equation}
Therefore, the calibration errors can be quantified in exactly the same
way as the waveform errors discussed in the previous section where
\begin{equation}
  (1 + \delta a(f)) \exp [i \delta \phi(f)] = 
  \left(\frac{R(f)}{R_{c}(f)}\right) \, . 
\end{equation}
This is precisely the form in which calibration errors are expressed,
for example in \cite{Dietz:2006}.  

The timing errors for a given phase accuracy are as given in the
previous section.  However, as has already been emphasized, the
calibration errors are uncorrelated between instruments, so expressions
(\ref{eq:timing_systematic}-\ref{eq:loose_timing}) are directly
applicable. 

\subsection{Example: Binary Coalescence}
\label{sec:cbc_systematic}

\begin{figure}[ht]
\centering
\includegraphics[width=.49\textwidth]{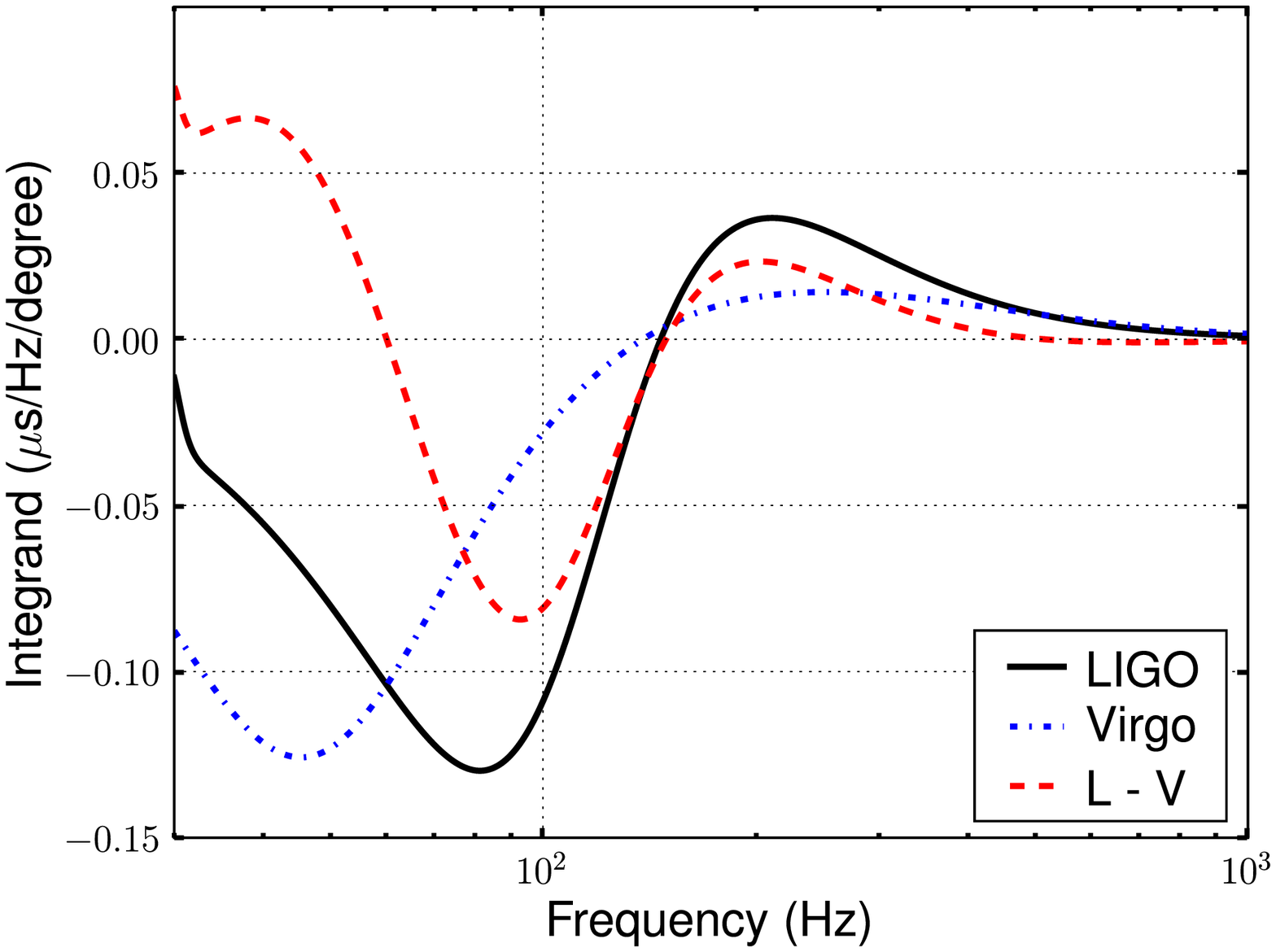}
\includegraphics[width=.49\textwidth]{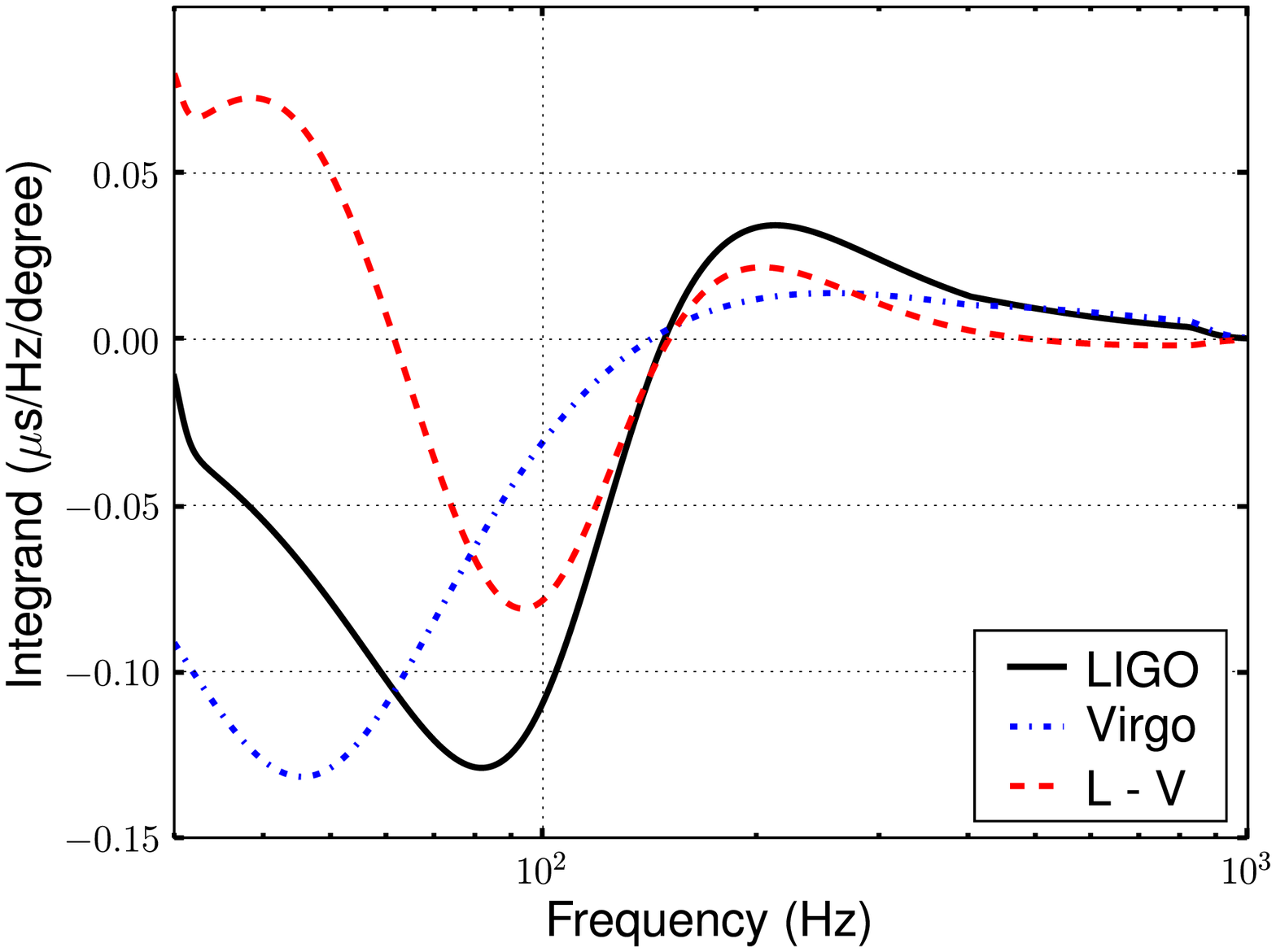}
\caption{The contribution that different frequencies make to the timing
offset given in equation (\ref{eq:timing_systematic}).  The left hand
plot is for the BNS system, the right hand plot is for the full $10-10
M_{\odot}$ coalescence waveform.  For both the LIGO and Virgo detectors,
frequencies far from the mean, but which still contribute to the signal
to noise ratio, contribute most significantly.  The different shapes are
directly dependent on the instrumental noise curves. The final curve
shows the contribution to the timing differential as a function of
frequency.}  
\label{fig:timing_integrand}
\end{figure}

Let us return once more to the coalescing binary waveforms and evaluate
the effect of waveform uncertainties on the timing.  In Figure
\ref{fig:timing_integrand} we plot the integrand of
(\ref{eq:timing_systematic}) for both the BNS and $10 - 10 M_{\odot}$
BBH signals.  The most significant contribution to timing errors arise
at frequencies above and below the mean frequency $\overline{f}$, but
where the detector still has good sensitivity.  Given a model of the
waveform error $\delta \phi(f)$, it can be integrated against the curves
in Figure \ref{fig:timing_integrand} to obtain the timing offset.

To calculate a worst case scenario for a given maximum phase offset
$\delta \phi_{\mathrm{max}}$, we take the absolute value of the
curves in Figure \ref{fig:timing_integrand} and integrate.  The
resultant errors for BNS are:%
\begin{equation}\label{eq:cal}
  |\delta t_{\mathrm{LIGO}}| \le \left[\frac{\delta
\phi_{\mathrm{max}}}{5^{\circ}} \right] 0.09 \mathrm{ms} 
  \, , \quad
  |\delta t_{\mathrm{Virgo}}| \le \left[\frac{\delta
\phi_{\mathrm{max}}}{5^{\circ}} \right] 0.06 \mathrm{ms} \, .
\end{equation}
These are limits on timing errors which would result in each detector
from a calibration error of this magnitude.  The errors would only be
achieved in the (rather unrealistic) circumstance where the phase error
was $+5^{\circ}$ up to the mean frequency and then $-5^{\circ}$ at all
frequencies above this.  

The use of an incorrect waveform family will introduce a correlated
timing error at the different sites.  In particular, it will not
introduce a timing systematic between the two LIGO sites, since the
detectors have the same noise power spectrum.  Figure
\ref{fig:timing_integrand} shows the relative timing offset between LIGO
and Virgo detectors.  It can again be integrated to obtain the maximal
timing effect of

\begin{equation}\label{eq:wf}
  |\delta t_{\mathrm{LIGO}} - \delta t_{\mathrm{Virgo}}| 
  \le \left[\frac{\delta \phi_{\mathrm{max}}}{5^{\circ}} \right] 0.05
 \mathrm{ms}
\end{equation}
For the BBH waveform, the $5^{\circ}$ errors are slightly larger at
$0.09$ ms for LIGO, $0.07$ ms for Virgo and $0.05$ ms between LIGO and
Virgo.   

It is illustrative to compare these systematic errors to the statistical
timing uncertainties derived in Section \ref{sec:timing}.  In the worst
case scenario, a calibration uncertainty in phase of $15^{\circ}$ would
introduce an error equivalent to the inherent timing uncertainty at
(single detector) SNR of 7.  Since the waveform errors will not affect
the timing between LIGO sites, they can be effectively absorbed into the
Virgo timing accuracy, in which case $20^{\circ}$ waveform uncertainties
will produce comparable uncertainties to inherent timing errors at SNR
7.  

\begin{table}[t]
\center
\begin{tabular}{| c | c | c | c | c | c | c | }
\hline \hline
Single Det & Waveform & Calibration & \multicolumn{2}{c|}{Timing (ms)} & 
  Median 90\% \\
SNR  & Error ($^\circ$) & Error ($^\circ$) & LIGO & Virgo & 
 Localization ($\deg^{2}$) \\ 
\hline
\hline
 7 &  - &  - & 0.27 & 0.21 & 40 \\
 7 & 10 & 10 & 0.32 & 0.26 & 60 \\
 7 & 20 & 20 & 0.45 & 0.38 & 120 \\
10 &  - &  - & 0.18 & 0.13 & 17 \\
10 &  5 &  5 & 0.20 & 0.15 & 22 \\
10 & 10 & 10 & 0.25 & 0.20 & 36 \\
25 &  - &  - & 0.06 & 0.05 &  2 \\
25 &  5 &  5 & 0.16 & 0.11 &  6 \\
25 & 10 & 10 & 0.19 & 0.13 & 20 \\
\hline
\hline
\end{tabular}
\caption{Timing and localization accuracy for binary neutron star
systems in the LIGO-Virgo network for a range of single detector SNRs
and calibration and waveform uncertainties.  The intrinsic timing
uncertainties are added in quadrature with the calibration and waveform
uncertainties.  The localization area is calculated from timing
information, with the additional assumption that amplitude consistency
can break the reflection degeneracy.  For an ideally located source
(orthogonal to the plane of the detectors) the localization area is half
the median value.  The numbers are given for the initial/enhanced
network; advanced detectors would yield around a 20\% improvement in
timing and 40\% in localization for the same SNR.} 
\label{tab:bns_errors}
\end{table}

Since the waveform, calibration and statistical uncertainties in timing
are independent, it is natural to add them in quadrature.  For a signal
at SNR 7 and $10^{\circ}$ waveform and calibration errors, we obtain
$\sigma_{l} \le 0.32 \mathrm ms$ for LIGO and $\sigma_{v} \le 0.25
\mathrm{ms}$ for Virgo.  This leads to a $50\%$ degradation of the
localization accuracy, so that for ideally located sources the area of
the $90\%$ confidence localization ellipse increases to $30 \deg^{2}$,
and for an average source to $60 \deg^{2}$.  Table \ref{tab:bns_errors}
provides the localization accuracies for sample of SNRs and
waveform/calibration uncertainties.  

\section{Discussion}
\label{sec:discussion}

We have introduced a simple method to compute the timing accuracy in a
gravitational wave detector and, using this, derived an expression for
the localization ability for a network of detectors.  For an
elliptically polarized waveform, we obtain a timing accuracy of
$\sigma_{t} \approx 1/(2 \pi \rho \sigma_{f})$.   Thus, the timing
accuracy scales inversely with both the amplitude and ``effective
bandwidth'' of the signal.  For reference, at a signal to noise ratio of
7, the timing accuracy for low mass coalescing binary signals is around
$0.25$ ms.  This holds for both BNS and low mass BBH, although for the
latter it is only by considering the full coalescence waveform that this
accuracy can be achieved. 

For a given timing accuracy in a network of detectors, we have
calculated the accuracy with which the source can be localized on the
sky.  The localization accuracy depends upon the timing accuracy in each
of the detectors, the network geometry and the angle between the plane
of the detectors and the signal location.  A detector network with
widely separated detectors affords the best localization ability, and
signals which are normal to the plane of a three detector network are
localized with the greatest accuracy.  The 90\% localization ellipse for
an optimally oriented source, with timing accuracy $0.25$ ms, in the
LIGO--Virgo network has an area of $20 \deg^{2}$.  This is doubled for
the median location, and doubled again if the reflection degeneracy in
the plane of the detectors cannot be broken by other considerations,
such as amplitude consistency.  

The expressions for localization accuracy derived here could easily be
extended to a network of more that three detectors.  Although the
precise form of the localization distribution has not been calculated,
it would be similar in form to the three detector expression
(\ref{eq:3site_post}).  In concurrence with other works
(e.g.~\cite{Blair:2008zz}) this argues that additional detectors would
most improve localization efforts by providing a large baseline between
the new detector and existing LIGO--Virgo network, as well as breaking
the reflection symmetry by lying well away from the plane formed by the
LIGO--Virgo network. 

We have given detailed expressions for the effect of systematic errors
on localization.  The effects of waveform and calibration uncertainties
is almost identical in a single detector.  However, the waveform errors
will be correlated across the detectors in the network, while
calibration errors will not.  Due to the similar sensitivity curves of
the two LIGO detectors, waveform uncertainties will have little effect
on localization, while they will effect the LIGO--Virgo network.
Calibration errors will be independent at the different sites and
therefore might have a larger effect on localization.  We have obtained
a bound on the timing error due to calibration errors as $\widehat
\delta t \le (\delta \phi_{\mathrm{max}}/2 \pi \sigma_{f})$.  By
comparing with the inherent timing uncertainty, we see that calibration
uncertainty will not dominate provided $\delta \phi_{\mathrm{max}} \le
1/\rho$.  Note, however, that only for a very specific, and unlikely,
form of the calibration error will the timing offset be anything like
this large.  

In this paper, we have not considered the effect of the mismatch between
waveform and template parameters.  This will surely degrade the
localization accuracy which has been derived.  However, as for waveform
error, the similarity of the LIGO detectors' sensitivities means that
waveform errors will have produce negligible timing effect between them,
as has been observed in \cite{Aylott:2009ya}.  Furthermore, in
\cite{Beauville:2007kp,Acernese:2007zza}, it has been argued that the
effect of parameter uncertainties can be minimized by choosing an
appropriate reference time.  In the future, we plan to investigate the
effect that parameter mismatch has on timing accuracies, as well as
exploring in greater detail the effect of waveform and calibration
errors for multi-detector parameter estimation.

There are numerous simplifications and approximations which are made in
this paper.  While the basic results derived here are qualitatively
correct, the detailed expressions for sky localization ellipses will
surely be modified as these assumptions are relaxed.  For example, if
the components of the binary are spinning, then the orbital plane will
precess during the evolution, whence the waveform will not be
elliptically polarized.  However, since the precession will be slow
relative to the gravitational wave frequency, it is reasonable to expect
that the results will be similar in the case of spinning binaries.
Likewise, the inclusion of higher waveform harmonics
\cite{VanDenBroeck:2006qi, VanDenBroeck:2006qu} will increase the
effective bandwidth of the signals; however it will also require a
generalization of the techniques described here to correctly incorporate
these signals.

Finally, we note that the localization estimates derived here are based
solely on triangulation between sites.  Thus, we have neglected
significant correlations which must be present in the observed signal at
more than two sites.  For three sites, there is an amplitude consistency
requirement that arises from the fact that the gravitational waveform
has only two polarizations.  Amplitude consistency should, in many
cases, serve to break the reflection degeneracy that arises by
considering triangulation alone and, in obtaining our most optimistic
localization results, we have assumed this is the case.  It is unlikely
that amplitude consistency will further restrict the location.  However,
as has been emphasized by Searle \cite{Searle:2009}, for a network of
three or more detectors, there is also a phase consistency between the
observed waveform at the sites.  For an elliptically polarized waveform
observed in two sites, there is always an orientation and polarization
such that the observed phase difference is consistent with a given sky
location.  However, for three sites, this is no longer the case and
phase consistency can be used to further restrict possible locations.
Any phasing requirement will naturally give a timing accuracy inversely
proportional to the signal's mean frequency $\bar{f}$.  Thus, in the
case where the mean frequency is significantly larger than the
bandwidth, one obtains higher frequency oscillations (from phasing) on
top of the slower falloff (from timing alone).  This will lead to
improvements in localization, and would be interesting to investigate
further.  However, for coalescing binaries, the mean frequency and
bandwidth are comparable which suggests this will not be a significant
effect. 

\section*{Acknowledgements}

We would like to thank many people for interesting discussions on this
topic, in particular John Baker, Duncan Brown, Kipp Cannon, Lee
Lindblom, Larry Price, Bangalore Sathyaprakash, Anthony Searle and Chris
Van Den Broeck.  We thank Patrick Sutton for many detailed discussions
and comments on the paper draft, and Ray Frey for detailed comments on
the paper draft.  This research was made possible thanks to support from
the Royal Society. 

\section*{References} 

\bibliographystyle{iopart-num}
\bibliography{iulpapers,refs,ninja}

\end{document}